\documentclass[cup7b,7pt]{cupbook}
\usepackage{graphicx}
\usepackage{multicol}
\usepackage{float}
\usepackage{subfigure}
\usepackage{setspace}
\usepackage[english]{babel}
\usepackage{verbatim}
\usepackage{fancyhdr}
\floatstyle{ruled}
\newfloat{tab}{thp}{lop}   
\floatname{tab}{Program}
\title[Rome, Italy, 27--30 April 2009]
      {The coming of age of X-ray polarimetry}
\begin{document}
\singlespacing
\pagenumbering{arabic}
\author[Francesco Lazzarotto et al.]{\begin{small}Francesco Lazzarotto\end{small} $^{a}$ \begin{small}Sergio Fabiani \end{small} $^{a}$ \begin{small}Enrico Costa\end{small} $^{a}$ \begin{small}Fabio Muleri\end{small} $^{a}$\and\begin{small}Paolo Soffitta\end{small} $^{a}$ \begin{small}Sergio Di Cosimo\end{small} $^{a}$ \begin{small}Giuseppe Di Persio\end{small} $^{a}$ \begin{small}Alda Rubini\end{small} $^{a}$  \and \begin{small}Ronaldo Bellazzini\end{small} $^{b}$ \begin{small}Alessandro Brez\end{small} $^{b}$ \begin{small}Gloria Spandre\end{small} $^{b}$ \begin{small}Vincenzo Cotroneo\end{small} $^{c}$ \and \begin{small}Alberto Moretti\end{small} $^{c}$ \begin{small}Giovanni Pareschi\end{small} $^{c}$ \begin{small}Giampiero Tagliaferri\end{small} $^{c}$}
\chapter{Angular Resolution of a Photoelectric Polarimeter in the Focus of an Optical System}
\begin{scriptsize}$^{a}$ IASF - INAF Roma\end{scriptsize}, 
\begin{scriptsize}$^{b}$ INFN Pisa\end{scriptsize}, 
\begin{scriptsize}$^{c}$ INAF Brera\end{scriptsize}, 
\begin{scriptsize}email: francesco.lazzarotto@iasf-roma.inaf.it\end{scriptsize} \begin{scriptsize}web: http://bigfoot.iasf-roma.inaf.it/$\sim$agile/Polar/SPSdoc/index.html\end{scriptsize}
\abstract{The INFN and INAF Italian research institutes developed a space-borne X-Ray polarimetry experiment based on a X-Ray telescope, focussing the radiation on a Gas Pixel Detector (GPD). The instrument obtains the polarization angle of the absorbed photons from the direction of emission of the photoelectrons as visualized in the GPD. Here we will show how we compute the angular resolution of such an instrument.}
\section{Introduction}
The X-ray telescopes are based on the grazing angle principle. The radiation is reflected with small incidence angles on the surfaces of hyperboloid and paraboloid mirrors and then is focused. The GPD is a gas detector which is able to image the photoelectron tracks. The polarization is measured using the dependence of the photoelectric cross section from the photon polarization direction \cite[Costa 2001]{Costa0} \cite[Bellazzini 2007]{Bellazzini}. The photoelectron is emitted with more probability in the direction of the electric field of the photon. The track created by the photoelectron path, is drifted and amplified by the Gas Electron Multiplier (GEM) and collected on a fine sub-divided pixel detector. Using different mixtures of gas it is possible to properly select the energy band of the instrument in the range of about $1-30$ keV. This GPD has the capability to preserve the imaging while reaching a good sensitivity in polarization as well as in spectroscopic and timing measurements.
\begin{table}
\begin{tabular}{|l|ccc}
\hline 
{\textbf{Characteristic}} & {\textbf{Value}} & {\textbf{Unit}}\\
\hline 
optics energy band & 0.1-10 & keV\\
GPD energy band & 1-30 & keV\\
GPD Area  & $15 \cdot 15$ & $mm^{2}$\\
GPD height & 10  & mm\\
GPD transistors & $16.5 \cdot 10^{6}$ & n. \\
GPD pixels & $105600$ & n. \\
GPD pixel matrix & $300 \cdot 352$ & n. \\
\hline 
\end{tabular}
\label{tab1}
\caption{GPD characteristics}
\end{table}
\section{Resolution Calculation and related Simulation Software}
We studied a system composed by an X-Ray telescope and the GPD. We considered only the on-axis radiation. In this case an ideally perfect optical system can focus the radiation exactly in a single point on the detector, assuming that it has:\begin{itemize}\item Perfect quality reflective surfaces; \item Perfectly coaxial alignment of the mirrors;\item A detector with negligible thickness.\end{itemize}
\begin{figure}[htpb]
\begin{tabular}{cc}
\begin{minipage}[c]{.30\textwidth}
\includegraphics[scale=0.21]{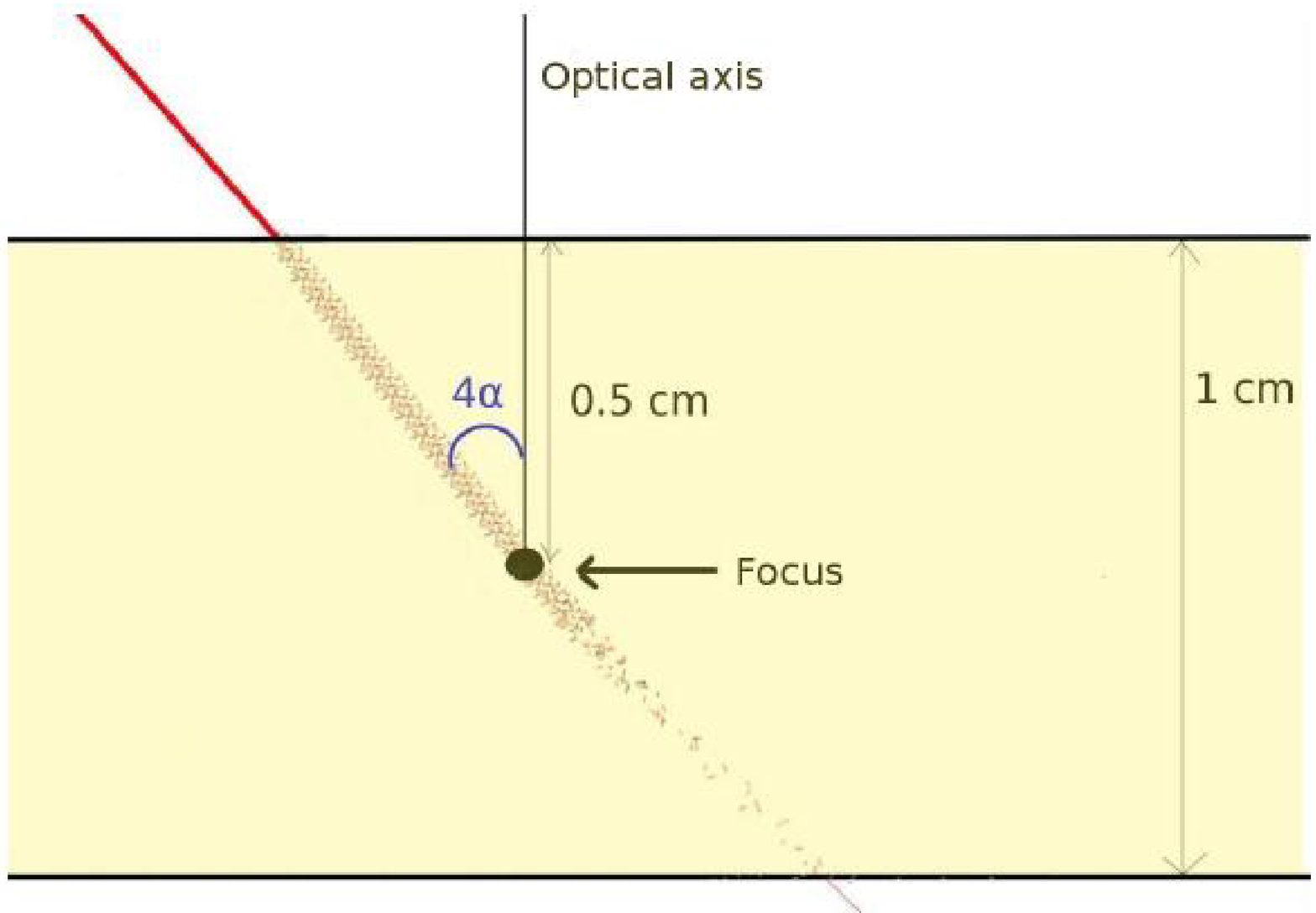}
\caption{Photon path in the GPD}
\end{minipage}
\hspace{2cm}
\begin{minipage}[c]{.30\textwidth}
\centering
\includegraphics[scale=0.11]{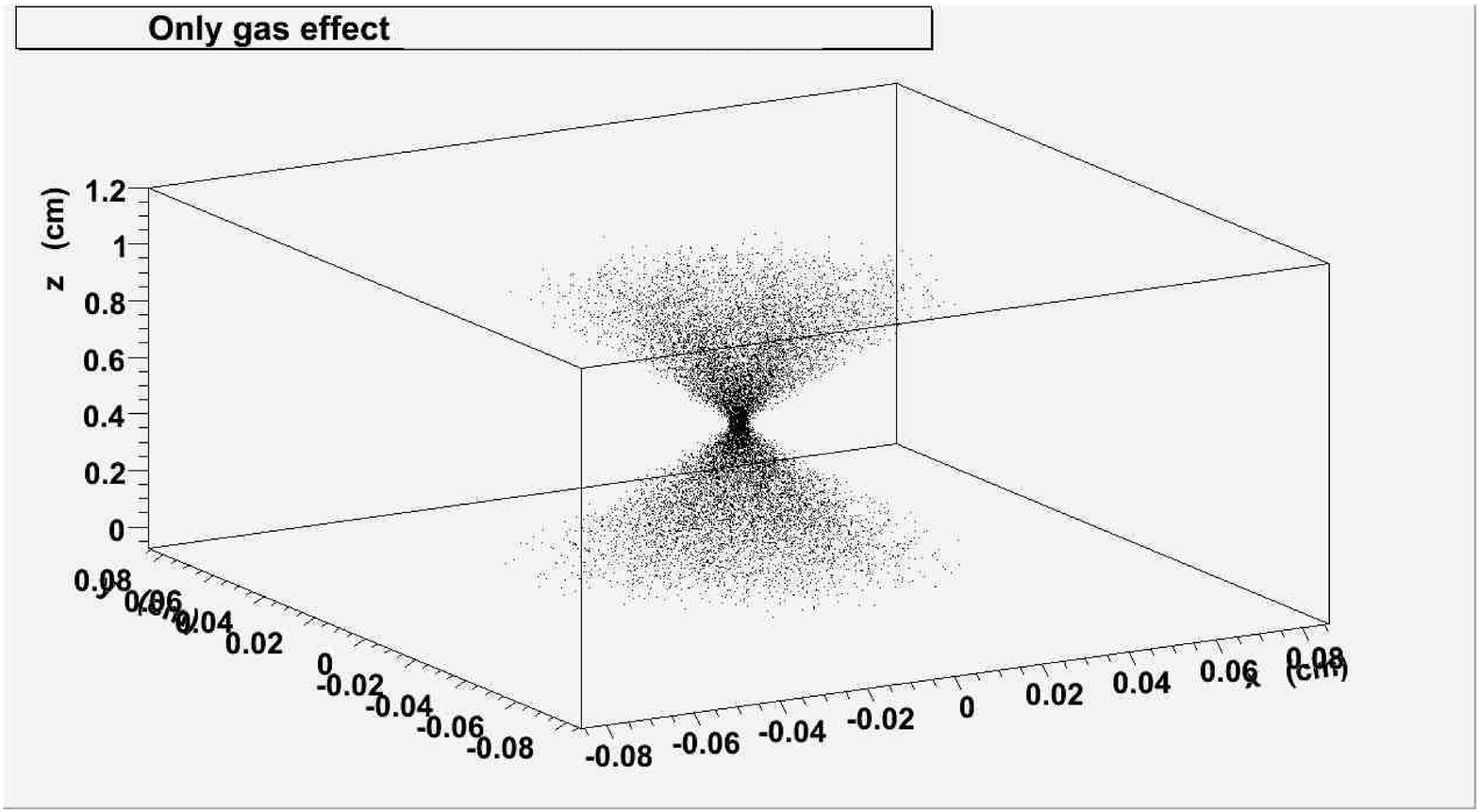}
\caption{Distribution of the absorbtion points in the gas cell of the GPD causing the gas blurring.}
\end{minipage}
 \end{tabular}
\end{figure}
\begin{figure}[!ht]
\centering
\includegraphics[height=4cm, width=11cm]{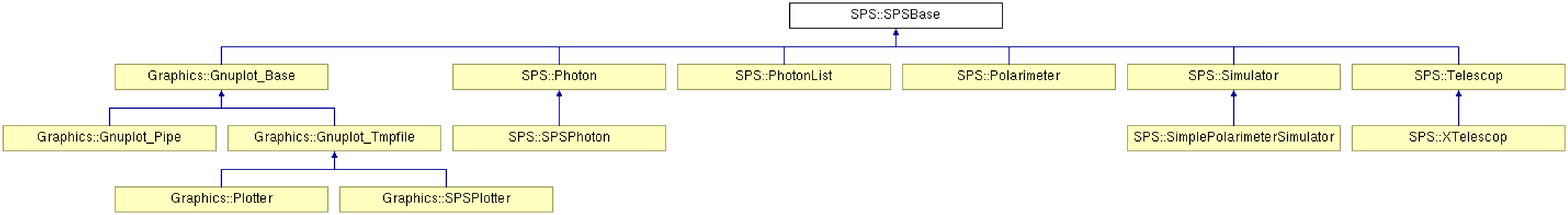}
\caption{Simulation software class tree}
\label{lazzarotto_f5}
\end{figure}
For the GPD the thickness of the gas cell is  not negligible: 1 cm. To express the real behavior of radiation intensity distribution, the Point Spread Function (PSF) is obtained taking into account: 
\begin{itemize}
\item Blurring introduced by imperfections of the optics \cite[Citterio, 1993]{Citterio93};
\item Blurring due to the approximations of the photons tracks reconstruction algorithm;
\item Blurring due to the radiation absorption in the gas.
\end{itemize}
We developed a simulation software based on montecarlo techniques to study the angular resolution of the instrument (see fig. \ref{lazzarotto_f5} \cite[Fabiani, 2008]{Fabiani1}). At this level the intrinsic resolution of the detector is neglected, the simulation program takes as input:
\begin{itemize}
\item The surface density of the incident radiation ($n. \ of \ photons  \cdot cm^{-2}$); 
\item The geometry and the effective area of the optical system;
\item The geometrical and physical characteristics of the gas detector. 
\end{itemize}
The program calculates the absorption point of the photons in the gas cell taking into account the effects of the optical aberrations and gas blurring. In output it produces graphics and statistics on the photon detection positions around the focal plane.
    \begin{figure}[!ht]
        \centering
        \begin{tabular}[!ht]{cc}
	\begin{minipage}[c]{.40\textwidth}
        \includegraphics[scale=0.4]{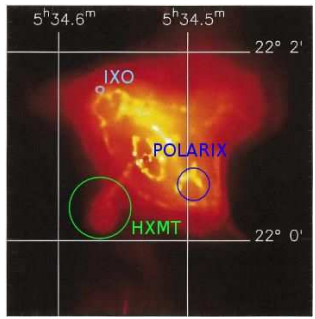} 
\caption{Qualitative representation of the resolution results on the image of the Crab PWN.}
\label{crabpwd}
\end{minipage}
\end{tabular}
\end{figure}
\section{Conclusion}
We report in the Table \ref{tab2} the angular resolution results expressed as the HPW [Half Power Width] of the radiation intensity on the detector plane for a simulation with a gas mixture composed by 70\% of DME and 30\% of He. In fig. \ref{crabpwd} the related error circles show that small missions as POLARIX and HXMT can be used to achieve the first results for the angular resolved X-Ray polarimetry. For instance it will be possible to measure the polarization of the main regions of extended sources such as the Pulsar Wind Nebulae. Whereas advanced missions as IXO will be able to investigate the thinner properties of such sources or to reach the resolution needed to resolve the knots in AGN jets.
\begin{table}
\begin{tabular}[!ht]{|l|c|c|c|c|}
{\textbf{\begin{small}Characteristic\end{small}}} & {\textbf{\begin{small}POLARIX\end{small}}} & {\textbf{\begin{small}HXMT\end{small}}}  & {\textbf{\begin{small}IXO\end{small}}} \\
\hline 
\begin{small}energy \end{small}& \begin{small}3 keV \end{small}& \begin{small}3 keV\end{small} & \begin{small}1.5 keV\end{small} & \\
\begin{small}HPW\end{small} \begin{small}gas + optics\end{small} & \begin{small}19.3 arcsec\end{small} & \begin{small}34.7 arcsec\end{small}  & \begin{small}6.6 arcsec\end{small}\\
\begin{small}HPW\end{small} \begin{small} only optics\end{small}  & \begin{small}14.7 arcsec\end{small} & \begin{small}23.2 arcsec\end{small}   & \begin{small}5.0 arcsec \end{small}\\
\begin{small}HPW\end{small} \begin{small} only gas\end{small} & \begin{small}10.0 arcsec\end{small} & \begin{small}19.5 arcsec\end{small} & \begin{small}3.0 arcsec \end{small} \\
\end{tabular}
\label{tab2}
\caption{Angular resolution, showing the different blurring contributions}
\end{table}

\clearpage
\end{document}